\documentclass[10pt]{article}
\usepackage[BOE]{express}
\usepackage{hyperref}
\usepackage{xfrac}

\graphicspath{{Images/}}

\begin{document}
\title{\textit{In vivo} laser Doppler holography of the human retina}

\author{L. Puyo,\authormark{1,*} M. Paques,\authormark{2,3} M. Fink,\authormark{1} J.-A. Sahel,\authormark{2,3}and M. Atlan\authormark{1}}

\address{\authormark{1}Institut Langevin. Centre National de la Recherche Scientifique (CNRS). Paris Sciences \& Lettres (PSL Research University). \'Ecole Sup\'erieure de Physique et de Chimie Industrielles (ESPCI Paris) - 1 rue Jussieu. 75005 Paris France\\
\authormark{2}Institut de la Vision, INSERM UMR-S 968. CNRS UMR 7210. UPMC. 17 rue Moreau, 75012 Paris France\\
\authormark{3}Centre d'investigation clinique (CIC) Centre Hospitalier National d'Ophtalmologie des Quinze-Vingts. INSERM. 28 rue de Charenton, 75012 Paris France}

\email{\authormark{*}gl.puyo@gmail.com} 



\begin{abstract}
The eye offers a unique opportunity for non-invasive exploration of cardiovascular diseases. Optical angiography in the retina requires sensitive measurements, which hinders conventional full-field laser Doppler imaging schemes. To overcome this limitation, we used digital holography to perform laser Doppler perfusion imaging of human retina with near-infrared light. Two imaging channels with a slow and a fast CMOS camera were used simultaneously for real-time narrowband measurements, and offline wideband measurements, respectively. The beat frequency spectrum of optical interferograms recorded with the fast (up to $75 \, \rm kHz$ ) CMOS camera was analyzed by short-time Fourier transformation. Power Doppler images drawn from the Doppler power spectrum density qualitatively revealed blood flow in retinal vessels over 512 $\times$ 512 pixels covering 2.4 $\times$ 2.4 mm$^2$ on the retina with a temporal resolution down to $1.6 \, \rm ms$. The sensitivity to lateral motion as well as the requirements in terms of sampling frequency are discussed.
\end{abstract}


\bibliographystyle{osajnl}

\section{Introduction}
The implication of vascular disorders has been questioned for major diseases affecting the retina such as diabetic retinopathy~\cite{Hardarson2012}, age-related macular deneneration~\cite{Pemp2008} and glaucoma~\cite{Cherecheanu2013}. Monitoring retinal blood flow appears crucial to understand the pathophysiology of these diseases. Most state of the art non-invasive instruments used for ocular blood flow assessment rely on speckle contrast analysis or laser Doppler methods. Both of these approaches use the intensity fluctuations of coherent light caused by moving objects in order to measure blood flow~\cite{Briers1995}.
Nowadays, Optical Coherence Tomography Angiography (OCT-A) has emerged as a powerful vascular imaging technique~\cite{Kashani2017}. OCT-A makes use of the speckle variations in the OCT signal caused by moving scatterers to generate an angiographic contrast by calculating the local variance of speckle realizations over a few repeated measurements~\cite{Wang2010a, Jia2012}. OCT-A instruments can map the retinal micro-vascular network with a micrometer axial resolution~\cite{Choi2013} and can be used to measure metrics relevant to the development of diabetic retinopathy such as the size and distribution of capillaries and the extent of the foveal avascular zone~\cite{Jia2015}. However, as it is necessary to reconstruct a full volume to obtain en-face images, the temporal resolution for swept-source OCT-A is of the order of a second, which is too slow for blood flow monitoring during a single cardiac cycle. 
Doppler OCT is a technique based on OCT that measures the Doppler frequency~\cite{Chen1997, Izatt1997} or phase~\cite{ZhaoChen2000} shift caused by the flow. The phase shift can be calculated between two consecutive measurements to extract a dynamic signal and reveal blood flow. However large flows causing a modulation higher than $2 \pi$ rad cannot be unambiguously calculated~\cite{Leitgeb2014}. Another limitation is that it cannot detect flow in vessels perpendicular to the incident OCT beam and only the axial component of the flow can be easily retrieved.
Laser speckle flowgraphy is another technique based on speckle contrast whose purpose is to image retinal blood flow quantitatively, although in arbitrary units~\cite{Fujii1994, Sugiyama2010}. En-face images of the retina are formed on the sensor and several images are combined to extract an angiographic contrast; this allows for full-field blood flow measurements with a high temporal resolution. Indeed a commercialized ophthalmic instrument based on this technology (LSFG NAVI, Softcare Ltd, Fukuoka, Japan) is able to image pulsatile flow in real-time with a temporal resolution of $33 \, \rm ms$. In contrast with OCT-A, laser speckle flowgraphy has no depth sectioning ability so the influence of choroidal flow is unclear. Inter-eye comparisons are not possible but intraindividual measurements are highly reproducible~\cite{Sugiyama2010}, which makes it an appropriate instrument to investigate ocular blood flow changes upon physiological or pharmacological stimuli.
%
Laser Doppler flowmetry is based on measuring the Doppler broadening of coherent light induced by moving red blood cells in vascularized tissues~\cite{RivaLasser1992, Leahy1999, RivaGeiserPetrig2009, RajanVarghese2009}. The frequency shift is measured by means of interference between Doppler shifted light and non shifted light. This method allows monitoring of blood flow only in a single area, but it has been implemented on scanning systems to monitor blood flow over a full-field~\cite{Michelson1996}. Blood flow and mean velocity in relatives units can be derived from calculations based on the first moments of the power spectrum density. As for laser speckle flowgraphy, interindividual comparisons are difficult whereas intraindividual measurements are reproducible~\cite{Schmetterer2007}. Laser Doppler flowmetry was used in a direct image detection scheme~\cite{Serov2002} but this method is too slow for retinal imaging due ocular safety limits.

Lately, holographic detection schemes such as full-field swept-source OCT~\cite{SpahrHillmann2015}, off-axis full-field time-domain OCT~\cite{Sudkamp16} or off-axis holographic line-field en-face OCT~\cite{Fechtig2015} have emerged. These techniques allow parallelized imaging and have been demonstrated for \textit{in vivo} human retinal imaging; they have been made possible by technological progress made by the camera industry which now allows for higher acquisition speed, and they seem to have opened the way for new possibilities especially in the realm of offline numerical processing~\cite{Hillmann2016, Ginner18}.
The laser Doppler holography technique we present here is conceptually close to laser Doppler flowmetry, but like the aforementioned methods it is derived from holography which allows for full-field imaging. In laser Doppler holography, the optical field serving as non-Doppler shifted light is a separate reference beam beating against the Doppler shifted optical field backscattered by the retina. This allows to increase as much as needed the power of the reference field impinging on the sensor and thus be able to work with very low exposure time. Interferograms are recorded with a high throughput camera and wideband measurements of the beat frequency of digitally reconstructed holograms are performed. The angiographic contrast is drawn from the Doppler spectral broadening of light backscattered by the retina and blood flow changes during cardiac cycles are revealed using a short-time Fourier transform analysis. We report here on the implementation of laser Doppler holographic perfusion imaging in the human retina. In contrast with previous attempts on rodent eyes~\cite{Pellizzari2016}, the holographic configuration is on-axis in order to maximize the field of view.

\section{Methods}

\subsection{Optical setup}
\begin{figure}[]
\centering
\includegraphics[width = 0.85\linewidth]{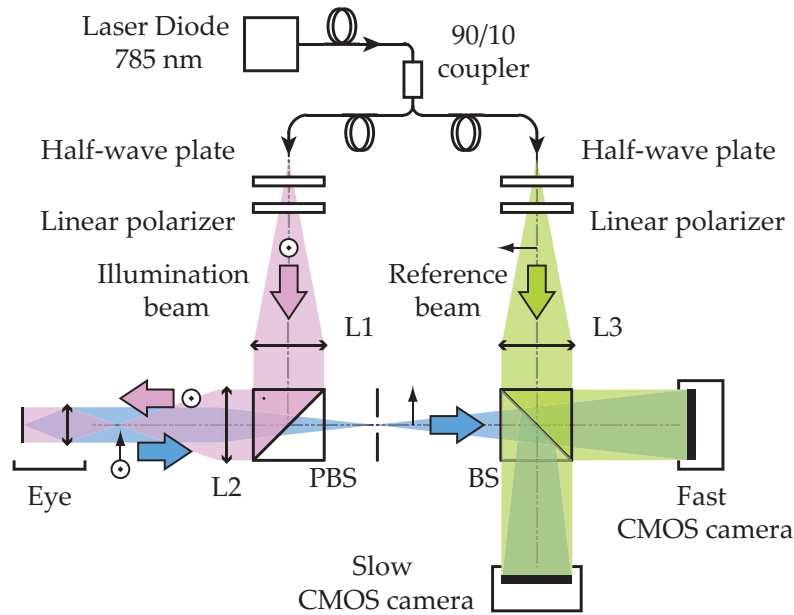}
\caption{Optical setup. L1, L2 and L3 are converging lenses. PBS: Polarizing Beam Splitter. BS: Beam Splitter. The light source is a single wavelength laser diode (SWL-7513-H-P, Newport). The 90\% output of the fiber coupler is used for the object arm. The Doppler broadened light backscattered by the retina is combined with the reference field and interferograms are recorded on the two imaging channels to study the Doppler beat frequency. The data from the fast CMOS camera is processed offline while the 80 Hz camera is used for real-time monitoring.
}
\label{fig_Setup}
\end{figure}

The optical setup designed and used for this study, shown in Fig.~\ref{fig_Setup}, consists of a fiber Mach-Zehnder optical interferometer. The light source used for the experiments is a $45 \, \rm  mW$, single-mode, fiber diode laser (Newport SWL-7513-H-P) at wavelength $\lambda = 785$ nm. The retina is illuminated with a $1.5 \, \rm  mW$ constant exposure over 2.4 $\times$ 2.4 mm$^2$ area. Informed consent was obtained for each subject and experimental procedures adhered to the tenets of the Declaration of Helsinki. The laser beam is focused in the front focal plane of the eye so that the light is collimated inside the eye and illuminates the retina on an extended area. This irradiation level is compliant with the exposure levels of the international standard for ophthalmic instruments ISO 15004-2:2007.

\begin{figure}[]
\centering
\includegraphics[width = 0.9\linewidth]{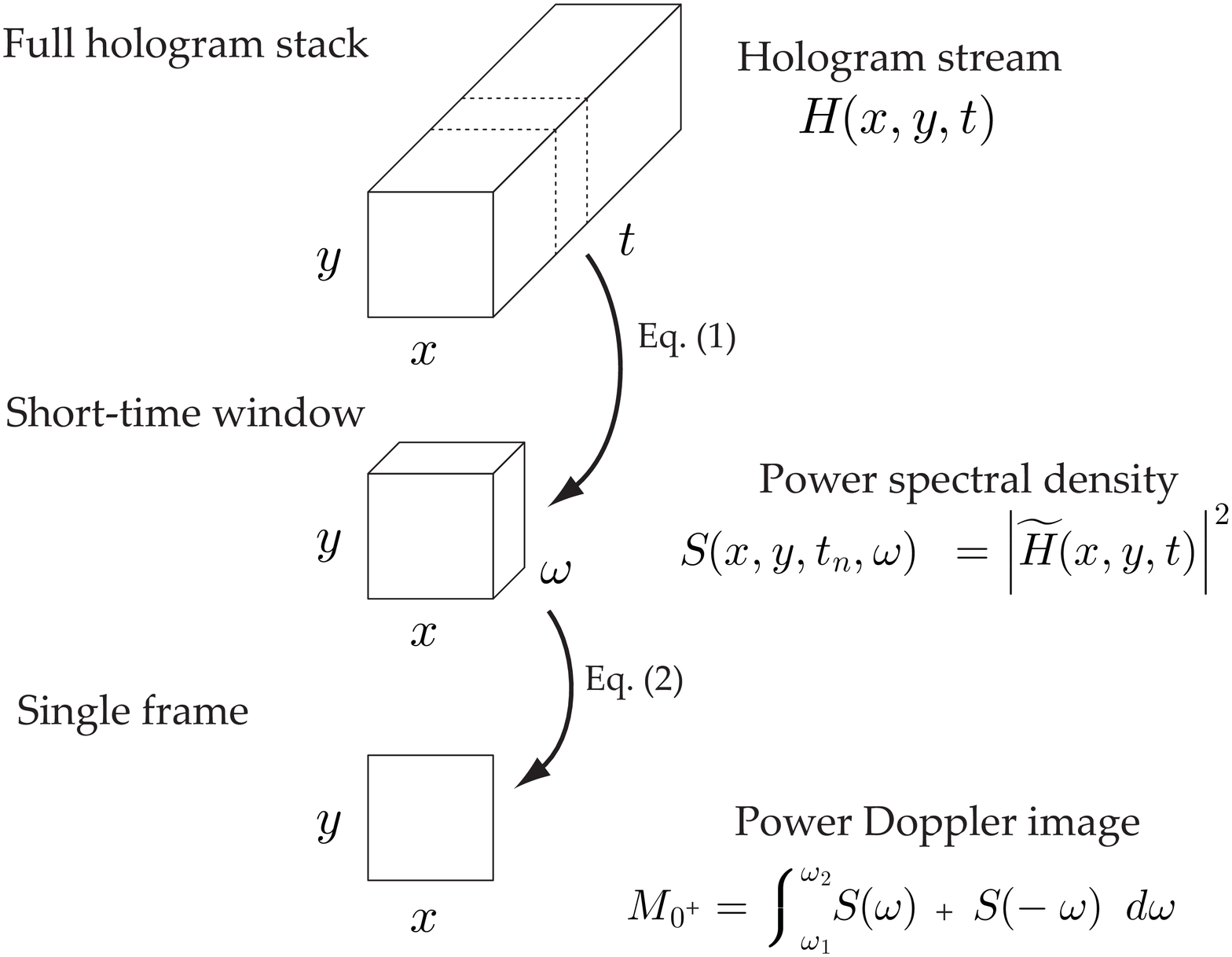}
\caption{Short-Time Fourier Transform analysis. A 3D sliding window of consecutive holograms is moved along the full hologram stack; for each window, the Doppler Power Spectral Density (DPSD) is estimated from the squared magnitude of the Fourier Transform in Eq. \eqref{eq:eq_PSD}. A single image with blood flow contrast is obtained by integrating the DPSD over $[\omega_{\rm 1}, \omega_{\rm 2}]$ in Eq. \eqref{eq:eq_Moments}, and the resulting image $M_{0^{+}}$ is referred to as power Doppler image. Examples of non-averaged power Doppler images are shown in Fig.~\ref{fig_Pulsatility_2}, and examples of averaged power Doppler images are shown in Fig.~\ref{fig_MomentsAll}(c) and Fig.~\ref{fig_Pulsatility_1}(c).}
\label{fig_STFT_Flowchart}
\end{figure}

A Polarizing Beam Splitter (PBS) cube is used in the object arm to illuminate the eye and collect the light backscattered by the retina. The PBS separates linear polarizations: it fully reflects the illumination beam and only the cross-polarized component of the light backscattered by the retina is transmitted to the camera. The reference beam $E_{\rm LO}$ is collimated and considered to be monochromatic and stable. The object and reference waves are combined using a non-polarizing beam splitter cube and they interfere on the sensor plane. The polarization of the reference wave is adjusted with a half-wave plate and a polarizer to optimize fringe contrast.

The interference pattern is recorded simultaneously on two imaging channels: a slow CMOS camera (Ximea - xiQ, frame rate $80 \, \rm Hz$, 2048 $\times$ 2048 pixels, 8 bit pixel depth, pixel size $5.5 \, \mu \rm m$) for real-time monitoring and a fast CMOS camera (Ametek - Phantom V2511, frame rate, up to $75 \, \rm kHz$ for a 512 $\times$ 512 format, quantum efficiency 40\%, 12-bit pixel depth, pixel size $28 \, \mu \rm m$) for offline processing.

\subsection{Laser Doppler measurements}
In laser Doppler measurement techniques, the Doppler effect is treated as a frequency shift of the optical field itself \cite{Albrecht2003}; here the optical field $E$ backscattered by the retina is considered to be the sum of the Doppler frequency shifted contributions. The interference pattern on the sensor is $I = \lvert E_{\rm LO} +E \rvert^{2}$. In the cross-beating terms, the optical field frequencies cancel out: the interferogram beat frequency spectrum carries the optical Doppler broadening. Doppler frequency shifts for biological tissues are small compared to the laser optical frequency $3.8 \times 10^{14} \, \rm Hz$ as the typical Doppler shift frequency spectra width for retinal vessels has been reported to be around 15-20 kHz~\cite{Riva1985}. However, as the spectral width corresponds to the Nyquist frequency of the fast camera, a major part of the Doppler broadening can be measured within the camera bandwidth in the interferogram beat frequency.

Acousto-optics modulators (not shown on the schematic) are used in conjuction with the slow camera to introduce a $40 \, \rm Hz$ frequency shift in order to perform stroboscopic and narrowband signal measurements in time-averaging conditions \cite{MagnainCastelBoucneau2014}. However due to eye motion the quality of the signal is limited and this channel is mostly used for alignment purposes. As both methods are based on holography, the slow camera gives indications on the quality of the holographic reconstruction that are also valid for the fast camera; moreover it is able to yield a contrast related to blood flow that allows visualizing the retinal area being imaged during the acquisition, which is not possible on the fast camera. The real-time hologram rendering is done with our software (Holovibes) that runs calculations on graphics processing units; for this purpose a personal computer equipped with a Nvidia GeForce GTX Titan graphics card is used.

The fast camera is used for wideband measurements of the Doppler spectral broadening. Interferograms from this camera are recorded on its on-board memory and the data are processed offline using Matlab. Interferograms are recorded on the $512 \times 512$ pixel sensor with an acquisition rate of $f_{\rm S}=39 \, \rm kHz$ (except in section \ref{section_FrequencyRole} where experiments are made to explore the effect of the sampling frequency), and a frame exposure time of $25.6 \, \mu{\rm s}$. Image rendering of complex-valued holograms $H(x,y)$ is performed using the angular spectrum propagation of the recorded interferograms $I(x,y)$, appropriate for near-field reconstructions~\cite{Goodman2005}:

\begin{eqnarray}\label{eq_AngularSpectrumDiffractionIntegral}
H(x,y,z) \approx \iint \hat{I}(k_x,k_y)  \exp{\left(i k_z z\right)} \exp{\left(ik_x x + ik_y y \right)} {\rm d} k_x {\rm d} k_y  
\end{eqnarray}
where $\hat{I}(k_x,k_y)$ is the two-dimensional Fourier transform of the interferogram $I(x,y)$. The optical axis is $z$, and $k_x$, $k_y$, and $k_z = \sqrt{k^2 - k_x^2 - k_y^2}$ are the wave vectors projections along $x$, $y$, and $z$, respectively. The lateral field of view of the reconstructed holograms of 512 $\times$ 512 pixels extends over 2.4 $\times$ 2.4 mm$^2$ on the retina.\\

\begin{figure}[]
\centering
\includegraphics[width = 0.85\linewidth]{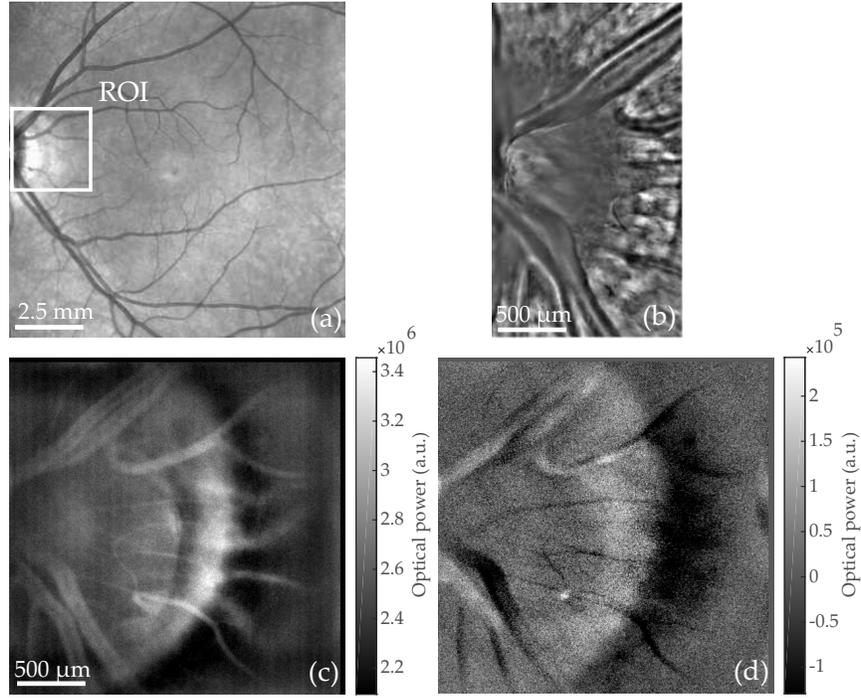}
\caption{Optic Nerve Head (ONH) region imaged with commercialized instruments and Laser Doppler holography. (a) Scanning laser ophthalmoscope (Spectralis, Heidelberg). (b) Adaptive optics flood illumination (rtx1, Imagine Eyes). (c) Power Doppler images $M_{0^{+}}$ calculated from holograms recorded at $f_{\rm S}=39 \, \rm kHz$; $S(\omega)$ is integrated over $[f_{\rm 1}, f_{\rm 2}]$ = 4-19.5 kHz. Multiple power Doppler images $M_{0^{+}}$ are averaged over a total time of $80 \, \rm ms$ (see \textcolor{blue}{Visualization 1} for blood flow movie). The peri-papillary crescent is visible on the edge of the ONH. (d) Asymmetry of the DPSD $M_{0^{-}}$ (averaged over the same period of time) illustrating the resultant flow direction with respect to the optical axis.}
\label{fig_MomentsAll}
\end{figure}

\subsection{Data processing}
Data processing consists of measuring the local optical fluctuations of the holograms recorded with the high-speed CMOS camera. A Short-Time Fourier Transform (STFT) method is used to analyze the changes in blood flow over time through the changes in the Doppler broadened spectrum. The variations over time of the Doppler Power Spectrum Density (DPSD) are resolved with a sliding short time window constituted of $j_{\rm win}$ consecutive holograms $H$. Given a sampling frequency $f_{\rm S}$, the duration of the short-time window is thus:
\begin{equation} \label{eq:eq_SigmaWin}
\sigma_{\rm win}= \frac{j_{\rm win}}{f_{\rm S}}
\end{equation}

For each short-time window, a single image with a blood flow contrast is formed using the process illustrated in Fig.~\ref{fig_STFT_Flowchart}. First, the short-time window is fast Fourier transformed along the temporal dimension, then the power spectral density is estimated from the squared magnitude of the Fourier transform:

\begin{equation} \label{eq:eq_PSD}
S(x,y,t_{n},\omega) = \left| \int_{t_{n}}^{t_{n}+ \sigma_{\rm win}} H(x,y,\tau) \exp{\left(-i \omega \tau\right)} \, {\rm d}\tau \right|^2 
\end{equation}

Finally the quantity $M_{0^{\pm}}$ is calculated from the integration of the DPSD:
\begin{equation}\label{eq:eq_Moments}
M_{0^{\pm}}(x,y,t_{n}) = \int_{\omega_{1}}^{\omega_{2}} S(x,y,t_{n},\omega) \: \pm  S(x,y,t_{n},-\omega) \  {\rm d} \omega 
\end{equation}
$M_{0^{+}}$ is the zeroth moment of the bandpass filtered DPSD; it is a single image corresponding to the area under the curve of the DPSD over the frequency interval $[f_{\rm 1}, f_{\rm 2}] = (2\pi)^{-1} \times [\omega_{\rm 1}, \omega_{\rm 2}]$. $M_{0^{-}}$ images reveal the asymmetry of the DPSD and are discussed in detail in section \ref{subsection_DopplerAsymmetry}. Temporal frequencies in this article are referred to indifferently as $f$ or $\omega = 2\pi f$. $M_{0^{+}}$ images are referred to as Power Doppler images (PDIs) and reveal the local blood flow. The frequency band $[f_{\rm 1}, f_{\rm 2}]$ is typically a high-pass filter (i.e. $f_{\rm 2} = {f_{\rm S}}/{2}$). The lower boundary $f_{\rm 1}$ is set between $4 \, \rm kHz$ and $7 \, \rm kHz$ in order to filter off the contributions that are not due to pulsatile flow in the vessels. The effect of the frequency band $[f_{\rm 1}, f_{\rm 2}]$ is explored in subsection \ref{subsection_Pyramide_Images}.
Once the $M_{0^{+}}$ image has been calculated, the sliding window is moved along the temporal dimension of the hologram stack. In this way, a sequence of PDIs revealing the blood flow for each moment of the cardiac cycle is made.

\section{Results}
\subsection{Blood flow contrast}

\begin{figure}[]
\centering
\includegraphics[width = 0.85\linewidth]{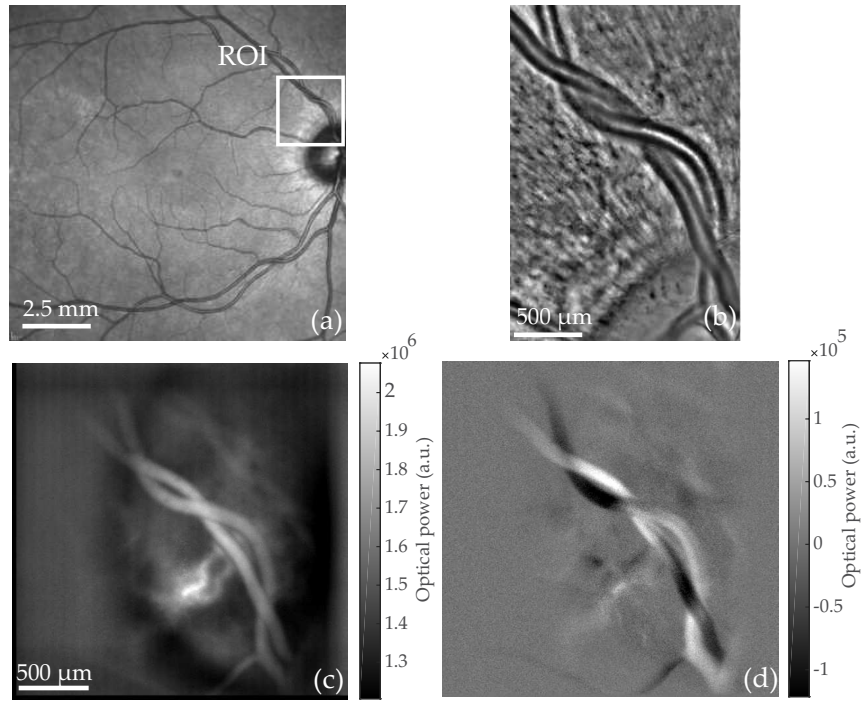}
\caption{An artery intertwined with a vein is imaged with commercialized instruments and Laser Doppler holography.(a) Scanning laser ophthalmoscope (Spectralis, Heidelberg). (b) Adaptive optics flood illumination (rtx1, Imagine Eyes). (c) Power Doppler images $M_{0^{+}}$ calculated from holograms recorded at $f_{\rm S}=39 \, \rm kHz$; $S(\omega)$ is integrated over $[f_{\rm 1}, f_{\rm 2}]$ = 7-19.5 kHz. Multiple power Doppler images $M_{0^{+}}$ are averaged over a total time of $0.66 \, \rm s$ (see \textcolor{blue}{Visualization 2} for blood flow movie). (d) Asymmetry of the DPSD $M_{0^{-}}$ (averaged over the same period of time) illustrating the resultant flow direction with respect to the optical axis.
}\label{fig_Pulsatility_1}
\end{figure}

Output images of the STFT analysis $M_{0^{+}}$ are displayed in Fig.~\ref{fig_MomentsAll}(c) and Fig.~\ref{fig_Pulsatility_1}(c) and associated visualizations linked in corresponding captions. These images reveal the tissue perfusion occurring during the duration of the short-time window $\sigma_{\rm win}$. Indeed the Doppler frequency shift is greater when the velocity of the scattering particle is greater. When integrating the whole high-pass filtered DPSD, the pixels where the Doppler broadening is the greatest have the brightest intensity on $M_{0^{+}}$ images. Thus areas of the retina where the local flow is greater are directly revealed in the intensity of PDIs.
 
As a result, the most noticeable features of the PDIs are the retinal vessels, due to the considerable flow of red blood cells moving at high speed. Blood vessels are clearly visible on PDIs and give a stronger signal than background tissues because of the larger flow in the vessels. Other features such as the nerve fiber layer are visible on reflectance images obtained with an adaptive optics flood illumination instrument (rtx1, Imagine Eyes, Orsay, France) in Fig.~\ref{fig_MomentsAll}(b) and in Fig.~\ref{fig_Pulsatility_1}(b), but these features are not visible on $M_{0^{+}}$ images.\\

\begin{figure}[]
\centering
\includegraphics[width = 0.85\linewidth]{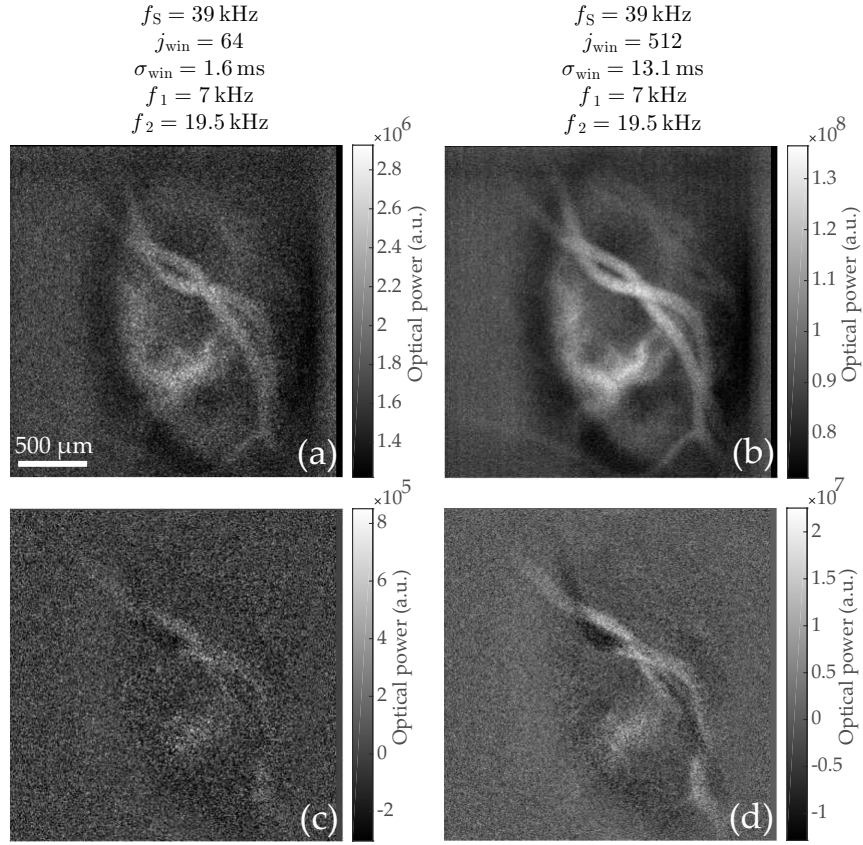}
\caption{Comparison in Signal to Noise Ratio (SNR) of power Doppler images $M_{0^{+}}$ and $M_{0^{-}}$ calculated with $j_{\rm win} = 64$ and 512 holograms (i.e. $\sigma_{\rm win}=1.6 \, \rm ms$ and $13.1 \, \rm ms$, respectively). The four images are non-averaged power Doppler images calculated from raw holograms acquired at $39 \, \rm kHz$:
(a) $M_{0^{+}}$, $j_{\rm win} = 64$.
(b) $M_{0^{+}}$, $j_{\rm win} = 512$.
(c) $M_{0^{-}}$, $j_{\rm win} = 64$.
(d) $M_{0^{-}}$, $j_{\rm win} = 512$.
The SNR dramatically improves with the duration of the short-time window. 
}\label{fig_Pulsatility_2}
\end{figure}

\begin{figure}[]
\centering
\includegraphics[width = \linewidth]{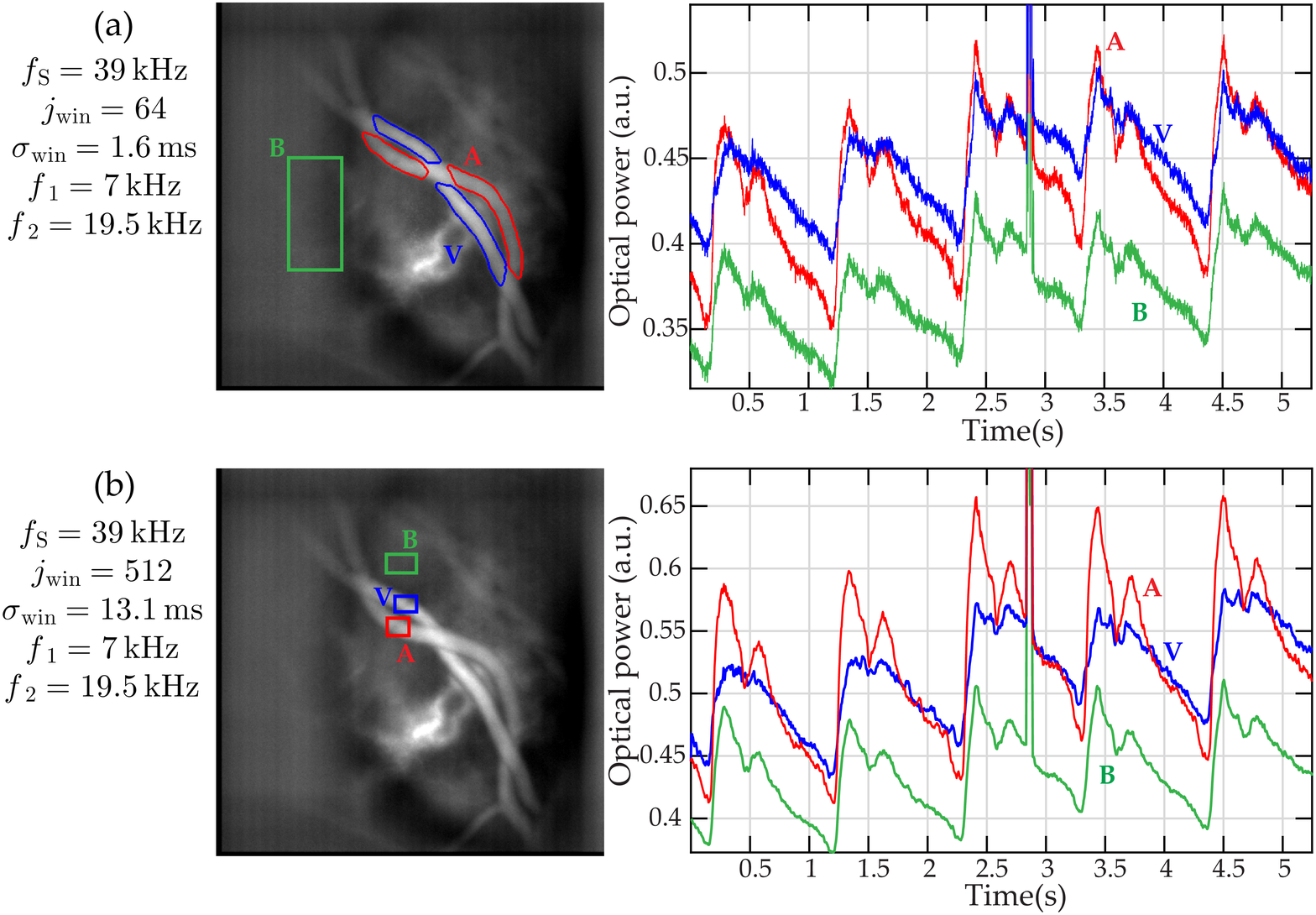}
\caption{Comparison of pulsatile flow measurements in a vein and an artery for two STFT analysis performed on the same dataset with (a) $j_{\rm win} = 64$ and (b)  $j_{\rm win} = 512$ holograms. For each analysis, a time averaged power Doppler images $M_{0^{+}}$ is displayed on the left hand side and shows a red, blue and green ROI which mark an artery (A), a vein (V), and the background tissues (B), respectively. These regions are used to spatially average the power Doppler signal for the plots on the right hand side. Although noisier, the pulsatility signal is already well visible when using $j_{\rm win} = 64$.
}\label{fig_Pulsatility_3}
\end{figure}

\subsection{STFT parameter $\sigma_{\rm win}$ }
The chosen width of the short-time window $\sigma_{\rm win}$ results from a trade-off between temporal resolution and Signal-to-Noise Ratio (SNR). Integrating the signal over a longer time period improves the SNR insofar as the amplitude of lateral displacements stays limited during the duration of the short-time window. For excessively long short-time windows, involuntary eye movements limit the improvement in SNR. More importantly, as two events happening during the same short-time window cannot be separated, the sliding window duration $\sigma_{\rm win}$ defines the temporal resolution of the instrument. In Fig.~\ref{fig_Pulsatility_2} and Fig.~\ref{fig_Pulsatility_3}, a comparison is made to evaluate the importance of $\sigma_{\rm win}$ on the SNR. The STFT analysis was carried out two times on the same dataset by using different parameters with an approximate tenfold increase in temporal resolution; $j_{\rm win}=64$ ($\sigma_{\rm win}=1.6 \, \rm ms$) in the first case and $j_{\rm win}=512$ ($\sigma_{\rm win}=13.1 \, \rm ms$) in the second case. As can be observed on individual PDIs in Fig.~\ref{fig_Pulsatility_2}, the SNR greatly improves when increasing $\sigma_{\rm win}$. Nonetheless the vasculature is already visible with $\sigma_{\rm win}=1.6 \, \rm ms$. Besides when looking at pulsatile flow, the shape of the power Doppler signal with $j_{\rm win}=64$ can show cardiac cycle signal the with a lower SNR than for $j_{\rm win}=64$.

For the STFT analysis conducted with $\sigma_{\rm win}=13.1 \, \rm ms$, a smaller ROI is required to spatially average the results to have a better SNR which illustrates as the SNR in the case where $j_{\rm win} = 64$ is lower.
Finally, it is interesting to notice that when averaging power Doppler images made with a STFT analysis performed with $j_{\rm win}=64$ and $j_{\rm win}=512$, the averaged image in the end are virtually the same as long as the total time on which PDIs are averaged is the same. This results are illustrated in images visible in Fig.~\ref{fig_Pulsatility_3}: in this case, 800 PDIs with $j_{\rm win}=64$ and 100 PDIs with $j_{\rm win}=512$ over the same period of time, and the two averaged PDIs show the same features with an equal SNR.

\subsection{Pulsatile flow}
The temporal resolution of the instrument is short enough to observe the changes in blood flow within cardiac cycles. When a heartbeat occurs, the increased speed and concentration of particles in vessels causes momentarily a larger Doppler spectral broadening of the optical field scattered in these structures, which is revealed by the STFT analysis. In Fig.~\ref{fig_Pulsatility_3}, the changes in blood flow throughout cardiac cycles can be observed in a retinal artery and vein. The power Doppler signal is spatially averaged in regions of interest corresponding to an artery, a vein, and background tissues; PDIs have been registered to compensate for lateral eye movements using~\cite{Guizar2008}. During systole (when the cardiac muscles contract and blood pressure is maximum) and diastole (when the cardiac muscles relaxe and the blood pressure is the lowest) the intensity of both vessels vary with the flow pulsatility. The artery gives a stronger signal than the vein during systole, because of the higher speed of circulating red blood cells which is in agreement with results reported in the literature \cite{White2003}. The artery also shows a larger modulation depth of the flow than the vein.

Interestingly, for each cardiac cycle, a first peak is reached at the time where the perfusion is maximum, and is followed by a second one of lower amplitude. This second peak corresponds to the dicrotic notch which is thought to due to the stretch and recoil of the aorta~\cite{Sabbah1978}. This waveform has previously been observed in various blood flow monitoring systems such as laser Doppler spectroscopy~\cite{Stern1977}, Doppler sonography~\cite{Polak2015}, Doppler OCT~\cite{Doblhoff2014}, or adaptive optics scanning laser ophthalmoscopy~\cite{ZhongPetrig2008}. It should be noted that the background tissue also exhibits a flow behavior that we assume to be due to unresolved retinal or choroidal capillaries. \\

\subsection{Laser Doppler spectral asymmetry} \label{subsection_DopplerAsymmetry}
\begin{figure}[]
\centering
\includegraphics[width = 0.75\linewidth]{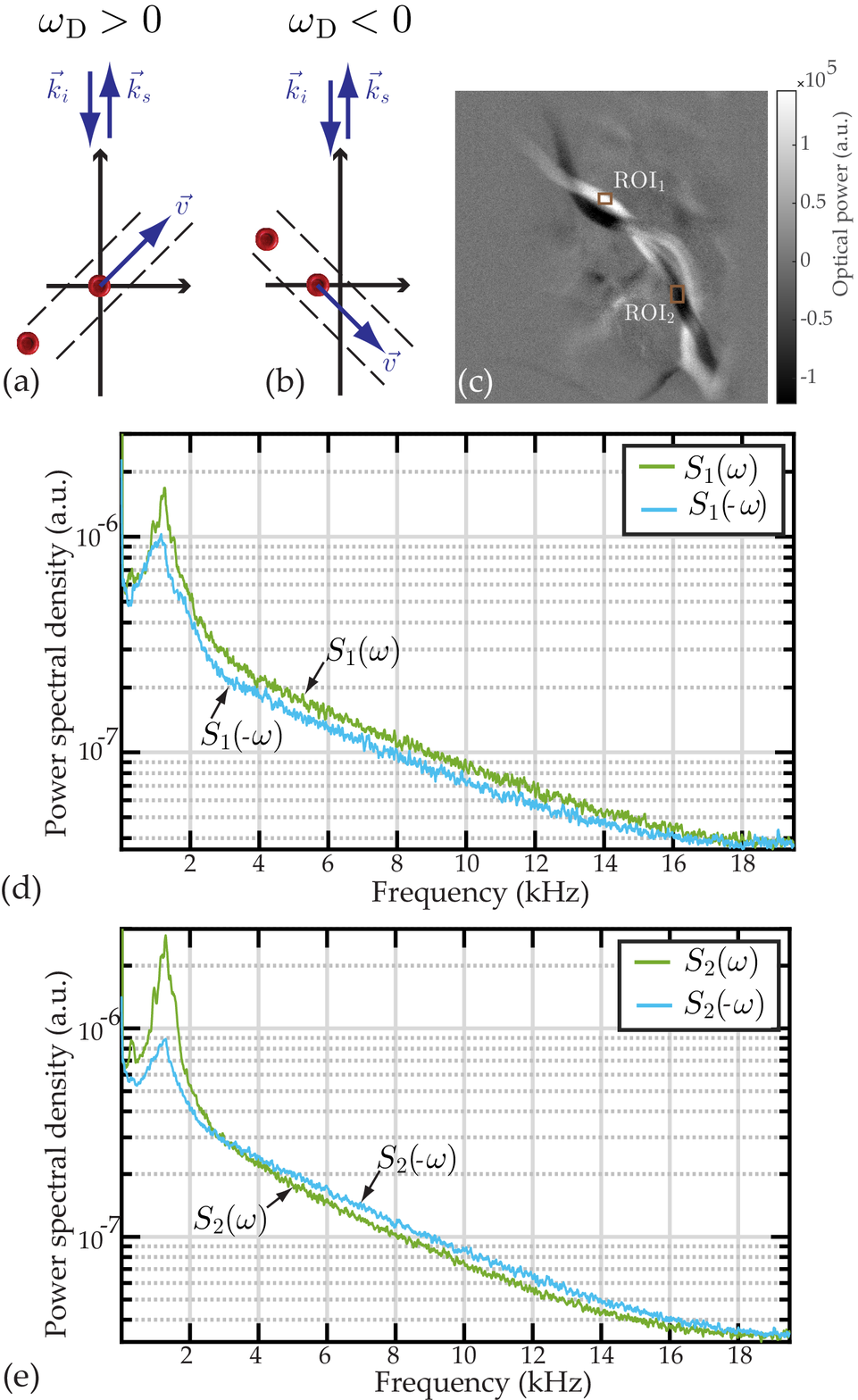}
\caption{Cases of spectral asymmetry between the positive and negative parts of the DPSD calculated according to Eq. \eqref{eq:eq_PSD}.
(a, b) Schematic illustrating the sign of the Doppler frequency shift for the direct backscatter (cf. Eq. \eqref{eq:eq_DopplerEff}).
(c) $M_{0^{-}}$ showing the two regions where the spectra are calculated.
(d) Negative and positive part of the DPSD measured in the ROI 1.
(e) Negative and positive part of the DPSD measured in the ROI 2.
For the range of frequencies 4-19.5 kHz used to calculate PDIs, $S_{\rm 1}(\omega) >  S_{\rm 1}(-\omega)$ and on the contrary $S_{\rm 2}(\omega) <  S_{\rm 2}(-\omega)$ depending on the vessel geometry.
}\label{fig_Pulsatility_4}
\end{figure}

In the full-field CMOS Doppler imaging scheme proposed by Serov et al.~\cite{Serov2002}, the light backscattered by the sample was not combined with a reference beam but was self-interfering. With this method, the non Doppler shifted light played the role of reference field to perform heterodyne measurements. In this configuration, a negative or positive Doppler frequency shift generates the same intensity fluctuations on the sensor: the signal recorded is real so its spectrum is exactly symmetric and the sign of the frequency shift cannot be recovered. In our case, as the reconstructed holograms are complex-valued and the phase of the optical field is known, the spectrum is no longer symmetric and it becomes possible to distinguish positive and negative frequency shifts. It can be observed that the difference between the negative and positive part of the DPSD contains information about the resultant flow direction when calculating the following quantity:\\
\begin{equation}\label{eq:eq_Moments_Neg}
M_{0^{-}}(x,y,t_{n}) = \int_{\omega_{1}}^{\omega_{2}} S(x,y,t_{n},\omega) -  S(x,y,t_{n},-\omega) \;  {\rm d} \omega 
\end{equation}
The asymmetry of the spectrum is in agreement with the vessel geometry (cf. Fig.~\ref{fig_MomentsAll}(d) and Fig.~\ref{fig_Pulsatility_1}(d) considering the expected sign of the Doppler shift. Indeed the Doppler frequency shift $\omega_{\rm D}$ occurring when the field is scattered by a moving object is calculated using the following formula:
\begin{equation}\label{eq:eq_DopplerEff}
\omega_{\rm D} = (\vec{k}_{\rm s}-\vec{k}_{\rm i}).\vec{v}
\end{equation}
Where $\vec{v}$ is the velocity vector of the scattering particle and $\vec{k}_{\rm i}$ and $\vec{k}_{\rm s}$ are the incoming and scattered optical wave vectors, respectively. According to the direction of the particle with respect to the optical axis, the frequency shift can be positive or negative as schematized in Fig.~\ref{fig_Pulsatility_4}(a) and Fig.~\ref{fig_Pulsatility_4}(b).
$M_{0^{-}}$ thus reveals the predominant direction of the flow with respect to the optical axis. If the vessels are not perpendicular to the optical axis, the projection of the velocity vector on the optical axis is non-null and the sign of the DPSD asymmetry varies accordingly with vessel directions. This is well visible in Fig.~\ref{fig_Pulsatility_4}(d) and Fig.~\ref{fig_Pulsatility_4}(e). The power spectrum density has been calculated according to Eq.\eqref{eq:eq_PSD} in two regions that differ from the direction of the local flow. In the first case the vessel is going slightly upward with respect to the optical axis and $S(\omega) > S(-\omega)$ over the frequency band $[f_{\rm 1}, f_{\rm 2}]$ which is used to integrate DPSD when calculating the PDIs. In the second case, the vessel is going downward and $S(\omega) < S(-\omega)$ over the frequency band of interest. This discrepancy between negative and positive parts of the DPSD is imputable to the directly backscattered light. It can also be noticed that for lower frequencies (for example the peak occurring at about $\approx 1 \, \rm kHz$), this inversion cannot be observed in two different regions as these frequencies correspond to Doppler shifts induced by global movements.

\section{Role of the frequency band $[f_{\rm 1}, \, f_{\rm 2}]$ and sampling frequency $f_{\rm S}$} \label{section_FrequencyRole}
In this section, results from several measurements performed at the same retinal location using different sampling frequencies (10, 20, 39, 75 kHz) are exploited to investigate on the effect of the frequency band $[f_{\rm 1}, \, f_{\rm 2}]$ and sampling frequency $f_{\rm S}$ on the resulting power Doppler images $M_{0^{+}}$. 
All measurements were done with the Phantom V2511 camera; the exposure time of the camera was set to the maximum for each acquisition (i.e. 100, 50, 25.6 ,$13.3 \, \rm {\mu s}$, respectively). The retina of a young and healthy subject was illuminated with the same exposure for all acquisitions, only the power of the reference arm was adjusted using neutral densities so that the camera was filled with the same amount of light (near saturation of the brightest pixels). The region imaged in these acquisitions is centered on the optic disc. The field of view is close to 3 $\times$ 3 mm$^2$, slightly larger than for the results shown in previous Figs. because we used a different combination of lenses.

\subsection{Choice for $[f_{\rm 1}, \, f_{\rm 2}]$} \label{subsection_Pyramide_Images}
\begin{figure}[]
\centering
\includegraphics[width = \linewidth]{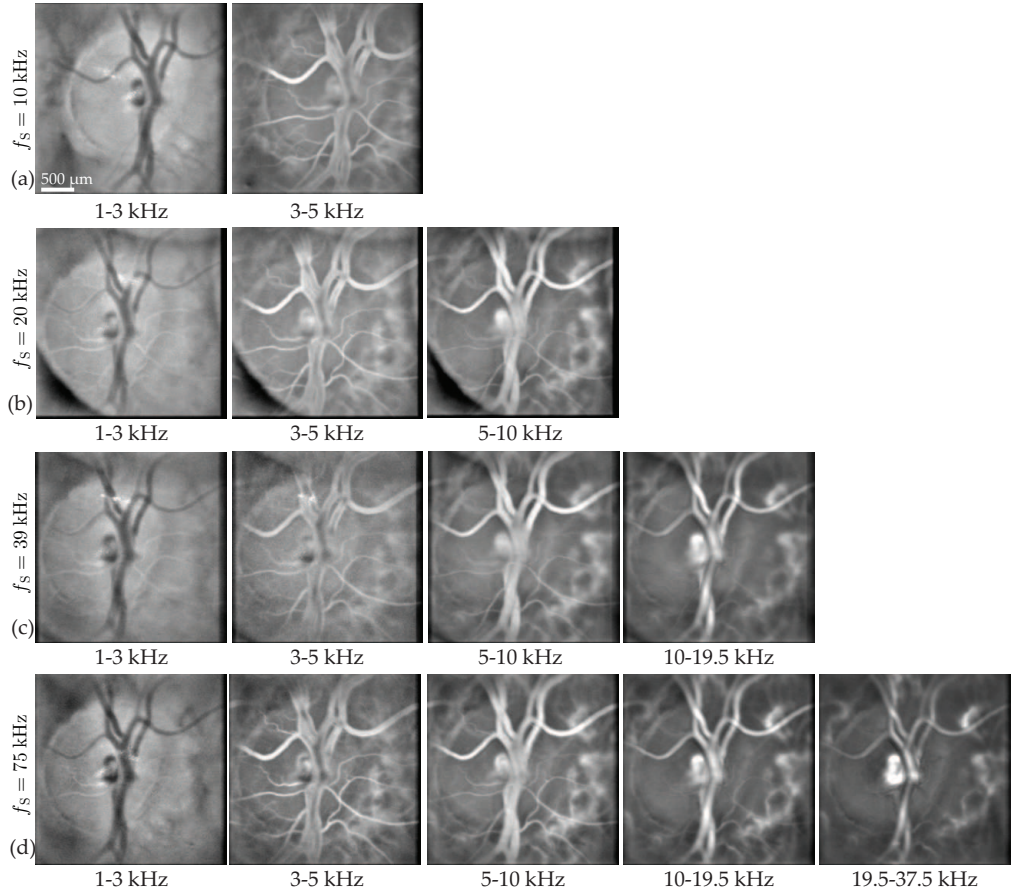} 
\caption{Multiple laser Doppler holography measurements of a same region are made with different sampling frequencies. For each of these sampling frequencies, PDIs $M_{0^{+}}$ are calculated for different frequency bands indicated below the images. (a) $f_{\rm S}=10 \, \rm kHz$. (b) $f_{\rm S}=20 \, \rm kHz$. (c) $f_{\rm S}=39 \, \rm kHz$. (d) $f_{\rm S}=75 \, \rm kHz$.
}
\label{fig_Pyramide_Images_10k_20k_39k_75k}
\end{figure}

The frequency band $[f_{\rm 1}, \, f_{\rm 2}]$ used to integrate the DPSD in Eq. \eqref{eq:eq_Moments} is chosen so as to filter off the DC and autocorrelation terms as well as the signals due to global eye movements while keeping the contribution of Doppler shifted light due to pulsatile flow in the vessels. The lower boundary is generally set between 4 and $7 \, \rm kHz$ depending on the retinal region being imaged while the upper boundary is set to the camera Nyquist frequency ${f_{\rm S}}/{2}$. In Fig.~\ref{fig_Pyramide_Images_10k_20k_39k_75k}, the displayed PDIs are calculated for different frequency bands for each sampling frequency. Each row of the Fig. corresponds to a sampling frequency, $f_{\rm S}=10$, 20, 39, and $75 \, \rm kHz$, for the rows (a), (b), (c), and (d), respectively. For each column, $M_{0^{+}}$ are calculated for several bandpass interval: $[f_{\rm 1}, \, f_{\rm 2}] =$ 1-3 kHz, 3-5 kHz, 5-10 kHz, 10-19.5 kHz and 19.5-37.5 kHz from left to right. Except for the acquisition with the highest sampling frequency, some intervals are above the Nyquist frequency of some sampling frequencies and consequently PDIs cannot be calculated for these frequencies. The visualization linked in caption Fig.~\ref{fig_Pyramide_Images_10k_20k_39k_75k} shows PDIs for a more precise range of frequencies in the case of the acquisition at $75 \, \rm kHz$.

It can be observed in Fig.~\ref{fig_Pyramide_Images_10k_20k_39k_75k} that for the lowest frequency band 1-3 kHz, the background tissue DPSD is greater than the vessels DPSD, so the vessels appear darker than the tissue on the PDIs. For the frequency band 3-5 kHz, the edges of the large retinal vessel appear brighter than the lumen, as the latter remains dark. That is because the flow is greater at the center of the vessels and the Doppler broadening occurring in these structures is still undersampled. However, for this frequency band, the Doppler broadening caused by the flow in smaller retinal vessels seems adequately sampled as they appear brightly. For frequencies superior to $10 \, \rm kHz$, the DPSD in the lumen seems adequately sampled. Thus integrating lower frequencies allows us to get the signal from smaller retinal vessels, although it also increases the contribution of global axial and lateral movements. It can be noticed that PDIs calculated using a certain frequency band look alike independently from the sampling frequency. For example PDIs calculated with the frequency band 1-3 kHz are virtually the same with the sampling frequencies $10 \, \rm kHz$ and $75 \, \rm kHz$. So in the former case, all the signal above $5 \, \rm kHz$ (Nyquist frequency for $f_{\rm S}=10 \, \rm kHz$) which is aliased has no visible effect on the resulting $M_{0^{+}}$ image.

This leads to the conclusion that to make a simple image of the vasculature, the aliased part of the DPSD does not make significant change on PDIs. It is also possible to conclude that the DPSD corresponding to the contributions of pulsatile flow in large vessels lies in frequency bands that are above $5-10 \, \rm kHz$ and up to  $37.5 \, \rm kHz$ depending on the size of the vessel.

\subsection{Sampling frequency $f_{\rm S}$} 

\begin{figure}[]
\centering
\includegraphics[width = \linewidth]{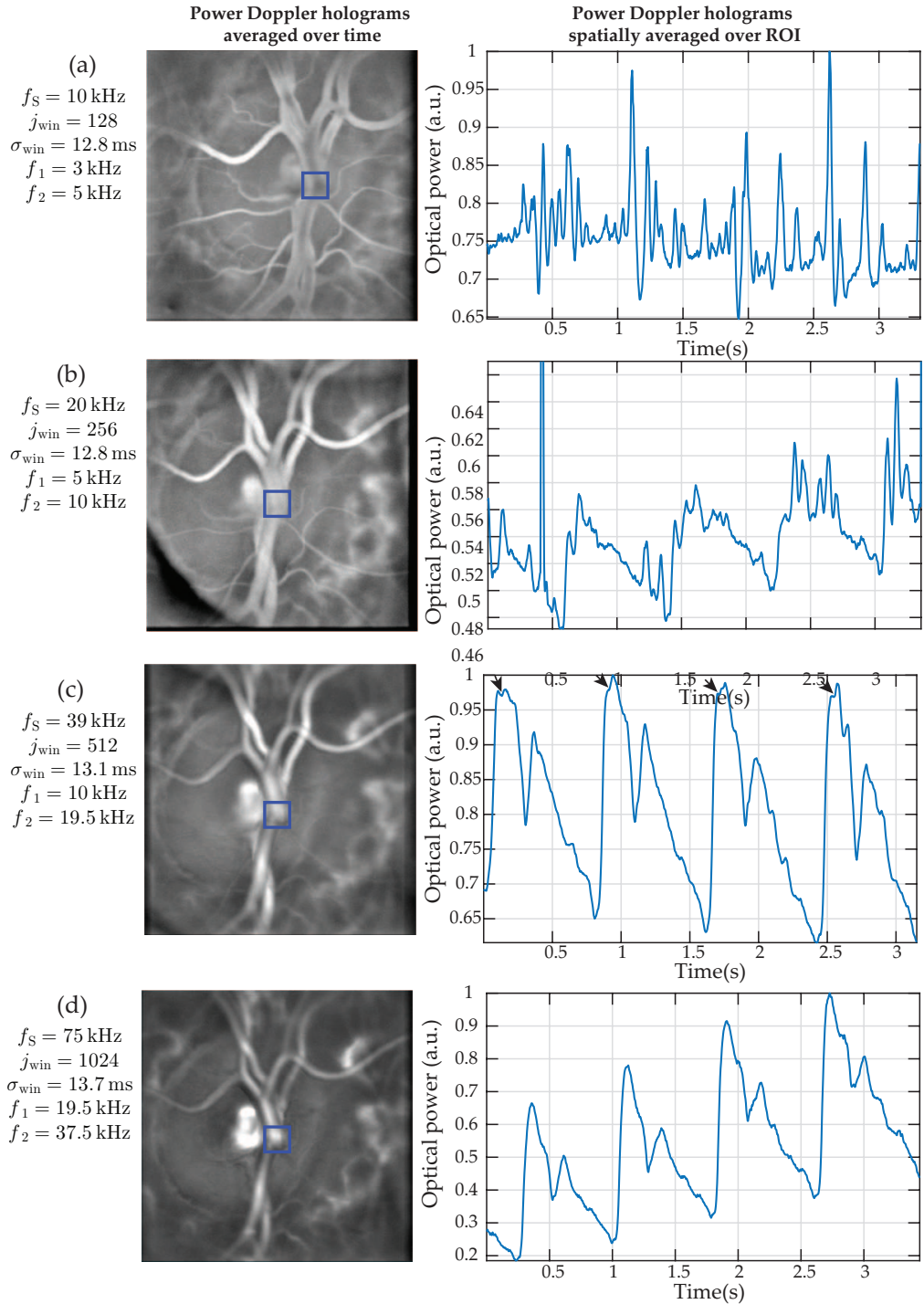} 
\caption{Requirements in terms of sampling frequency for laser Doppler holography in the central retinal artery. For each sampling frequency, power Doppler images are spatially averaged over the depicted ROIs on the left hand side and the result is displayed in the associated plot. The STFT parameters used for each acquisition are displayed on the left and have been chosen to have $\sigma_{\rm win}  \approx 13 \, \rm ms$ for all acquisitions. (a) $f_{\rm S}=10 \, \rm kHz$, (b) $f_{\rm S}=20 \, \rm kHz$, (c) $f_{\rm S}=39 \, \rm kHz$, (d) $f_{\rm S}=75 \, \rm kHz$. \textcolor{blue}{Visualization 3} shows PDIs as a function of the frequency for the acquisition with $f_{\rm S}=75 \, \rm kHz$.
}
\label{fig_Plots_10k_20k_39k_75k}
\end{figure}

In Fig.~\ref{fig_Plots_10k_20k_39k_75k}, the requirements in terms of sampling frequency to perform laser Doppler holography in the human retina are investigated.
For this purpose, both the vasculature images of the retina $M_{0^{+}}$ and the changes in power Doppler signal during the cardiac cycles are considered. 
STFT analysis are performed with $j_{\rm win }=128$, 256, 512, 1024 holograms for the acquisitions at $f_{\rm S}=10$, 20, 39, and $75 \, \rm kHz$ respectively. In this way the temporal resolutions of the STFT analyses are approximately the same for all acquisitions ($\sigma_{\rm win}  \approx 13 \, \rm ms$). Power Doppler images $M_{0^{+}}$ are calculated using the highest possible frequency bands according to the sampling frequency: $[f_{\rm 1}, \, f_{\rm 2}] =$ 3-5 kHz, 5-10 kHz, 10-19.5 and 19.5-37.5 kHz, for $f_{\rm S}=10$, 20, 39, and $75 \, \rm kHz$, respectively.
For each sampling frequency, an image calculated from PDIs averaged over time is displayed on the left, and the plots on the right correspond to the signal over time of $M_{0^{+}}$ averaged in the region indicated on the associated image. This ROI is centered on the arterial bifurcation of the central retinal artery where the flow is expected to be the greatest in the retina.

As observed in the previous subsection \ref{subsection_Pyramide_Images}, PDIs of the vasculature at lower frequencies show a weaker signal inside the vessel lumen because the Doppler broadening is undersampled. The curve for the lowest frequency band 3-5 kHz displayed on the right side on the Fig. does not show a pattern related to the cardiac cycle. However for the frequency band above (5-10 kHz), the typical curve of flow over time in the vessels is recognizable. Then for the two highest frequency bands, the difference is minor. Indeed when looking closely it is possible to see that around the maximum of the systolic peak for the 10-19.5 kHz plot, the curve is not well rounded and shows an abrupt decrease (indicated by the arrows) that is not visible for 19.5-37.5 kHz plots. We assume that a minor undersampling briefly occurs at this moment for large systolic flows. In Fig.~\ref{fig_Plots_10k_20k_39k_75k}(d), an overall increase over time of the power Doppler signal can be noticed: it is due to an involuntary eye movement which progressively shifts the pupil position.

Hence Fig.~\ref{fig_Plots_10k_20k_39k_75k} shows that the Doppler shifted light corresponding to the pulsatile flow in large vessels is located further along the power spectrum density. Considering these results, performing measurements with a sampling frequency above 30 to $40 \, \rm kHz$ seems to be a requirement for large retinal vessels. Otherwise, the part of the DPSD corresponding to the pulsatile flow is undersampled, aliased, and mixed with low frequencies of greater amplitude and is therefore not retrievable. However, for vessels of smaller size it is possible to use a smaller sampling frequency and have a properly sampled Doppler signal but there is a risk that this signal might be too weak due to the low number of circulating red cells especially compared to the Doppler shifts induced by global eye movements.

\section{Discussion and conclusion}
It is possible to record up to 259,329 frames in a 512 $\times$ 512 format on the on-board memory of the Phantom V2511 camera. This corresponds to approximately $6.6 \, \rm s$ of acquisition at $f_{\rm S}=39 \, \rm kHz$ or $3.5 \, \rm s$ of acquisition at $f_{\rm S}=75 \, \rm kHz$. We showed in the previous section that with $f_{\rm S}=39 \, \rm kHz$ the Doppler broadening caused by pulsatile flow seems somewhat undersampled only when imaging the central retinal artery during systole. As a result, when imaging other retinal regions where the flow is weaker, this sampling frequency yields negligible aliasing. Consequently, for this work the sampling frequency was usually set to $39 \, \rm kHz$ to be able to record longer acquisitions and maximize our chances to measure consecutive cardiac cycles unaffected by micro-saccades. Indeed, as with most ophthalmic instruments, the ability of laser Doppler holography to perform retinal imaging during micro-saccades is limited. The Doppler broadening induced by global movements is greater than the one induced by local pulsatile flows (for example the strong peak occurring at $2.8 \, \rm s$ in Fig.~\ref{fig_Pulsatility_3} is a micro-saccade). Consequently, during a micro-saccade, the intensity of the PDIs is predominated by global movements. However drifts and tremor do not prevent blood flow imaging, as can be observed in movies because slow movements generate low frequency Doppler shifts that are removed when high pass filtering the DPSD. It should be noted that this process also filters off the Doppler signal due to the flow in small retinal vessels.

The technique presented in this article makes use of the backscattered light that has been highly Doppler shifted. This part of the light is selected when high-pass filtering the DPSD in Eq.\eqref{eq:eq_PSD}. One of the assumptions of the theory used for laser Doppler flowmetry is that photons scattered by red blood cells (i.e. highly Doppler shifted photons) have been scattered by static diffusers before they impinge on erythrocytes~\cite{BonnerNossal1981, Bonner1990}. The result of this is a randomization of the wave vectors involved in the scalar product in Eq. \eqref{eq:eq_DopplerEff} yielding the Doppler frequency shift: the situation is comparable to red blood cells being illuminated from all directions. Thus, selecting highly Doppler shifted light implies selecting multiply scattered light. Yet multiply scattered light loses the original linear polarization of the incident beam because light is depolarized in the multiple scattering regime~\cite{Schmitt1992, RojasOchoa2004}. Consequently, it is possible to carry out laser Doppler holography with cross-polarized light because we work with the backscattered light that is highly Doppler shifted, thus multiply scattered and depolarized. For this reason we used a polarizing beam splitter cube which allows us to filter off the specular reflection occurring on the cornea. Indeed this reflection conserves the original linear polarization of the incident beam and is reflected by the polarizing beam splitter cube instead of being transmitted to the camera.

An important aspect of working with multiply scattered light is that the instrument is sensitive to lateral motion. For conventional blood flow monitoring methods based on Doppler approaches, the flow is assessed in a simple manner only in vessels parallel to the optical axis. The flow in vessels perpendicular to the optical axis cannot be assessed as the Doppler shift is minimum, and for vessels with intermediate inclinations, the geometry must be known to calculate the corresponding flow. In our case, because we select multiply scattered light, vessels with any inclinations with respect to the optical axis contribute to the Doppler broadening of the backscattered light. Consequently the instrument is sensitive to lateral motion and able to image blood vessels in en-face planes as can be seen in the PDIs shown in Figs.~\ref{fig_MomentsAll}, \ref{fig_Pulsatility_1}, and \ref{fig_Pyramide_Images_10k_20k_39k_75k}.

However in laser Doppler flowmetry, the retinal area where the blood flow is monitored is not a large blood vessel but tissues that are mainly vascularized by capillaries. In this medium the randomization of the light is a safer assumption because small vessels are oriented in all directions and are surrounded by a dense matrix of static diffusers. However in our case it can be argued that the multiple scattering assumption is less valid as the vessel walls and other structures around large vessels may not diffuse the incident beam sufficiently, and the limits of this assumption can be seen in the asymmetry of the DPSD $M_{0^{-}}$. Indeed in a situation of purely multiple scattered light there should not be any difference between the positive and negative parts of the spectrum despite the vessels geometry since the asymmetry exists because of the signed Doppler shifts of the directly backscattered light. However, as explained in section \ref{subsection_DopplerAsymmetry}, a mild asymmetry does exist in the DPSD depending on the vessel geometry (less than 10 to 15\% in the example shown). This asymmetry may cause a bias of flow sensitivity according to the vessels orientation with respect to the optical axis. However the impact seems limited and is mitigated by the process of averaging the positive and negative parts of the spectrum. Moreover in the parts of vessels where the flow direction is perpendicular to the optical axis (i.e. the areas where $M_{0^{-}}=0$), the flow sensitivity does not seem to be affected by the  local inclination of the vessel with respect to the optical axis. 

In conclusion, wideband laser Doppler holography can reveal retinal blood flow pulsatility in humans over a full-field with a temporal resolution of only $1.6 \, \rm ms$ while keeping the retinal exposure under $1.5 \, \rm mW$ at $785 \, \rm nm$. The angiographic contrast which is sensitive to local dynamics, including lateral motion, is drawn from the Doppler power spectrum density using a short-time Fourier transform analysis. Thanks to its high temporal resolution, laser Doppler holography could provide new insight in the dynamics of the vascular system as well as potential new ways to diagnose and follow-up major retinal or cardio-vascular diseases.

\section*{Funding}
This work was supported by LABEX WIFI (Laboratory of Excellence ANR-10-LABX-24) within the French Program Investments for the Future under Reference ANR-10-IDEX-0001-02 PSL, and European Research Council (ERC Synergy HELMHOLTZ, grant agreement \#610110). The Titan Xp used for this research was donated by the NVIDIA Corporation.

\section*{Acknowledgments}
The authors would like to thank S. Meimon, K. Grieve and J-P. Huignard for helpful discussions.

\section*{Disclosures}
The authors declare that there are no conflicts of interest related to this article.

\begin{thebibliography}{10}
\newcommand{\enquote}[1]{``#1''}

\bibitem{Hardarson2012}
S.~H. Hardarson and E.~Stef{\'a}nsson, \enquote{Retinal oxygen saturation is altered in diabetic retinopathy,} {\protect\JournalTitle{British Journal of Ophthalmology}} \textbf{96}, 560--563 (2012).

\bibitem{Pemp2008}
B.~Pemp and L.~Schmetterer, \enquote{Ocular blood flow in diabetes and age-related macular degeneration,} {\protect\JournalTitle{Canadian Journal of Ophthalmology/Journal Canadien d'Ophtalmologie}} \textbf{43}, 295--301
  (2008).

\bibitem{Cherecheanu2013}
A.~P. Cherecheanu, G.~Garhofer, D.~Schmidl, R.~Werkmeister, and L.~Schmetterer, \enquote{Ocular perfusion pressure and ocular blood flow in glaucoma,} {\protect\JournalTitle{Current Opinion in Pharmacology}} \textbf{13}, 36--42
  (2013).

\bibitem{Briers1995}
J.~A. Briers and S.~Webster, \enquote{Quasi real-time digital version of single-exposure speckle photography for full-field monitoring of velocity or flow fields,} {\protect\JournalTitle{Optics Comm.}} \textbf{116}, 36--42 (1995).

\bibitem{Kashani2017}
A.~H. Kashani, C.-L. Chen, J.~K. Gahm, F.~Zheng, G.~M. Richter, P.~J. Rosenfeld, Y.~Shi, and R.~K. Wang, \enquote{Optical coherence tomography angiography: A comprehensive review of current methods and clinical applications,} {\protect\JournalTitle{Progress in Retinal and Eye Research}}
  (2017).

\bibitem{Wang2010a}
R.~K. Wang, L.~An, P.~Francis, and D.~J. Wilson, \enquote{Depth-resolved imaging of capillary networks in retina and choroid using ultrahigh sensitive optical microangiography,} {\protect\JournalTitle{Optics Letters}}
  \textbf{35}, 1467--1469 (2010).

\bibitem{Jia2012}
Y.~Jia, J.~C. Morrison, J.~Tokayer, O.~Tan, L.~Lombardi, B.~Baumann, C.~D. Lu, W.~Choi, J.~G. Fujimoto, and D.~Huang, \enquote{Quantitative OCT angiography of optic nerve head blood flow,} {\protect\JournalTitle{Biomedical Optics Express}} \textbf{3}, 3127--3137 (2012).

\bibitem{Choi2013}
W.~Choi, K.~J. Mohler, B.~Potsaid, C.~D. Lu, J.~J. Liu, V.~Jayaraman, A.~E. Cable, J.~S. Duker, R.~Huber, and J.~G. Fujimoto, \enquote{Choriocapillaris and choroidal microvasculature imaging with ultrahigh speed OCT angiography,} {\protect\JournalTitle{PloS One}} \textbf{8}, e81499 (2013).

\bibitem{Jia2015}
Y.~Jia, S.~T. Bailey, T.~S. Hwang, S.~M. McClintic, S.~S. Gao, M.~E. Pennesi, C.~J. Flaxel, A.~K. Lauer, D.~J. Wilson, J.~Hornegger, J.~G. Fujimoto, and D. Huang, \enquote{Quantitative optical coherence tomography angiography of vascular abnormalities in the living human eye,} {\protect\JournalTitle{Proceedings of the National Academy of Sciences}} \textbf{112}, E2395--E2402 (2015).

\bibitem{Chen1997}
Z.~Chen, T.~E. Milner, S.~Srinivas, X.~Wang, A.~Malekafzali, M.~J. van Gemert, and J.~S. Nelson, \enquote{Non-invasive imaging of in vivo blood flow velocity using optical Doppler tomography,} {\protect\JournalTitle{Optics Letters}}
  \textbf{22}, 1119--1121 (1997).

\bibitem{Izatt1997}
J.~A. Izatt, M.~D. Kulkarni, S.~Yazdanfar, J.~K. Barton, and A.~J. Welch, \enquote{In vivo bidirectional color Doppler flow imaging of picoliter blood volumes using optical coherence tomography,} {\protect\JournalTitle{Optics Letters}} \textbf{22}, 1439--1441 (1997).

\bibitem{ZhaoChen2000}
Y.~Zhao, Z.~Chen, C.~Saxer, S.~Xiang, J.~F. de~Boer, and J.~S. Nelson, \enquote{Phase-resolved optical coherence tomography and optical Doppler tomography for imaging blood flow in human skin with fast scanning speed and high velocity sensitivity,} {\protect\JournalTitle{Optics Letters}}
  \textbf{25}, 114--116 (2000).

\bibitem{Leitgeb2014}
R.~A. Leitgeb, R.~M. Werkmeister, C.~Blatter, and L.~Schmetterer, \enquote{Doppler Optical Coherence Tomography,} {\protect\JournalTitle{Progress in Retinal and Eye Research}} \textbf{41}, 26--43 (2014).

\bibitem{Fujii1994}
H.~Fujii, \enquote{Visualisation of retinal blood flow by laser speckle flowgraphy,} {\protect\JournalTitle{Medical and Biological Engineering and Computing}} \textbf{32}, 302--304 (1994).

\bibitem{Sugiyama2010}
T.~Sugiyama, M.~Araie, C.~E. Riva, L.~Schmetterer, and S.~Orgul, \enquote{Use of laser speckle flowgraphy in ocular blood flow research,} {\protect\JournalTitle{Acta Ophthalmologica}} \textbf{88}, 723--729 (2010).

\bibitem{RivaLasser1992}
C.~Riva, S.~Harino, B.~Petrig, and R.~Shonat, \enquote{Laser Doppler flowmetry in the optic nerve,} {\protect\JournalTitle{Experimental Eye Research}} \textbf{55}, 499--506 (1992).

\bibitem{Leahy1999}
M.~Leahy, F.~De~Mul, G.~Nilsson, and R.~Maniewski, \enquote{Principles and practice of the laser-Doppler perfusion technique,} {\protect\JournalTitle{Technology and Health Care}} \textbf{7}, 143--162 (1999).

\bibitem{RivaGeiserPetrig2009}
C.~E. Riva, M.~Geiser, and B.~L. Petrig, \enquote{Ocular blood flow assessment using continuous laser Doppler flowmetry,} {\protect\JournalTitle{Acta Ophthalmologica}} \textbf{88}, 622--629 (2009).

\bibitem{RajanVarghese2009}
V.~Rajan, B.~Varghese, T.~G. van Leeuwen, and W.~Steenbergen, \enquote{Review of methodological developments in laser Doppler flowmetry,} {\protect\JournalTitle{Lasers in Medical Science}} \textbf{24}, 269--283
  (2009).

\bibitem{Michelson1996}
G.~Michelson, B.~Schmauss, M.~Langhans, J.~Harazny, and M.~Groh, \enquote{Principle, validity, and reliability of scanning laser Doppler flowmetry.} {\protect\JournalTitle{J. Glaucoma.}} \textbf{5}, 99--105 (1996).

\bibitem{Schmetterer2007}
L.~Schmetterer and G.~Garhofer, \enquote{How can blood flow be measured?} {\protect\JournalTitle{Survey of Ophthalmology}} \textbf{52}, S134--S138 (2007).

\bibitem{Serov2002}
A.~Serov, W.~Steenbergen, and F.~de~Mul, \enquote{Laser Doppler perfusion imaging with complementary metal oxide semiconductor image sensor,} {\protect\JournalTitle{Optics Letters}} \textbf{27}, 300--302 (2002).

\bibitem{SpahrHillmann2015}
H.~Spahr, D.~Hillmann, C.~Hain, C.~Pf{\"a}ffle, H.~Sudkamp, G.~Franke, and G.~H{\"u}ttmann, \enquote{Imaging pulse wave propagation in human retinal vessels using full-field swept-source optical coherence tomography,} {\protect\JournalTitle{Optics letters}} \textbf{40}, 4771--4774 (2015).

\bibitem{Sudkamp16}
H.~Sudkamp, P.~Koch, H.~Spahr, D.~Hillmann, G.~Franke, M.~M\"{u}nst, F.~Reinholz, R.~Birngruber, and G.~H\"{u}ttmann, \enquote{In-vivo retinal imaging with off-axis full-field time-domain optical coherence tomography,} {\protect\JournalTitle{Optics Letters}} \textbf{41}, 4987--4990 (2016).

\bibitem{Fechtig2015}
D.~J. Fechtig, B.~Grajciar, T.~Schmoll, C.~Blatter, R.~M. Werkmeister, W.~Drexler, and R.~A. Leitgeb, \enquote{Line-field parallel swept source MHz OCT for structural and functional retinal imaging,} {\protect\JournalTitle{Biomedical Optics Express}} \textbf{6}, 716--735 (2015).

\bibitem{Hillmann2016}
D.~Hillmann, H.~Spahr, C.~Hain, H.~Sudkamp, G.~Franke, C.~Pf{\"a}ffle, C.~Winter, and G.~H{\"u}ttmann, \enquote{Aberration-free volumetric high-speed imaging of in vivo retina,} {\protect\JournalTitle{Scientific Reports}} \textbf{6}, 35209 (2016).

\bibitem{Ginner18}
L.~Ginner, T.~Schmoll, A.~Kumar, M.~Salas, N.~Pricoupenko, L.~M. Wurster, and R.~A. Leitgeb, \enquote{Holographic line field en-face OCT with digital adaptive optics in the retina in vivo,} {\protect\JournalTitle{Biomed. Opt. Express}} \textbf{9}, 472--485 (2018).

\bibitem{Pellizzari2016}
M.~Pellizzari, M.~Simonutti, J.~Degardin, J.-A. Sahel, M.~Fink, M.~Paques, and M.~Atlan, \enquote{High speed optical holography of retinal blood flow,} {\protect\JournalTitle{Optics letters}} \textbf{41}, 3503--3506 (2016).

\bibitem{Albrecht2003}
H.~Albrecht, M.~Borys, N.~Damaschke, and C.~Tropea, \emph{Laser Doppler and Phase Doppler Measurement Techniques} (Springer, 2003).

\bibitem{Riva1985}
C.~E. Riva, J.~E. Grunwald, S.~H. Sinclair, and B.~Petrig, \enquote{Blood velocity and volumetric flow rate in human retinal vessels.} {\protect\JournalTitle{Investigative Ophthalmology \& Visual Science}}
  \textbf{26}, 1124--1132 (1985).

\bibitem{MagnainCastelBoucneau2014}
C.~Magnain, A.~Castel, T.~Boucneau, M.~Simonutti, I.~Ferezou, A.~Rancillac, T.~Vitalis, J.-A. Sahel, M.~Paques, and M.~Atlan, \enquote{Holographic laser Doppler imaging of microvascular blood flow,} {\protect\JournalTitle{JOSA A}}
  \textbf{31}, 2723--2735 (2014).

\bibitem{Goodman2005}
J.~W. Goodman, \emph{Introduction to Fourier Optics} (Roberts and Company Publishers, 2005).

\bibitem{Guizar2008}
M.~Guizar-Sicairos, S.~T. Thurman, and J.~R. Fienup, \enquote{Efficient subpixel image registration algorithms,} {\protect\JournalTitle{Optics Letters}} \textbf{33}, 156--158 (2008).

\bibitem{White2003}
B.~R. White, M.~C. Pierce, N.~Nassif, B.~Cense, B.~H. Park, G.~J. Tearney, B.~E. Bouma, T.~C. Chen, and J.~F. de~Boer, \enquote{In vivo dynamic human retinal blood flow imaging using ultra-high-speed spectral domain optical Doppler tomography,} {\protect\JournalTitle{Optics Express}} \textbf{11},
  3490--3497 (2003).

\bibitem{Sabbah1978}
H.~N. Sabbah and P.~D. Stein, \enquote{Valve origin of the aortic incisura,}
  {\protect\JournalTitle{American Journal of Cardiology}} \textbf{41}, 32--38
  (1978).

\bibitem{Stern1977}
M.~Stern, D.~Lappe, P.~Bowen, J.~Chimosky, G.~Holloway, H.~Keiser, and R.~Bowman, \enquote{Continuous measurement of tissue blood flow by laser-Doppler spectroscopy.} {\protect\JournalTitle{American Journal of Physiology}} \textbf{232}, H441--H448 (1977).

\bibitem{Polak2015}
J.~F. Polak, J.~M. Alessi-Chinetti, A.~R. Patel, and J.~M. Estes, \enquote{Association of common carotid artery Doppler-determined dicrotic notch velocity with the left ventricular ejection fraction,}
  {\protect\JournalTitle{Journal of Ultrasound in Medicine}} \textbf{34}, 461--467 (2015).

\bibitem{Doblhoff2014}
V.~Doblhoff-Dier, L.~Schmetterer, W.~Vilser, G.~Garh{\"o}fer, M.~Gr{\"o}schl, R.~A. Leitgeb, and R.~M. Werkmeister, \enquote{Measurement of the total retinal blood flow using dual beam Fourier-domain Doppler optical coherence tomography with orthogonal detection planes,}
  {\protect\JournalTitle{Biomedical Optics Express}} \textbf{5}, 630--642
  (2014).

\bibitem{ZhongPetrig2008}
Z.~Zhong, B.~L. Petrig, X.~Qi, and S.~A. Burns, \enquote{In vivo measurement of erythrocyte velocity and retinal blood flow using adaptive optics scanning laser ophthalmoscopy,} {\protect\JournalTitle{Optics Express}} \textbf{16}, 12746--12756 (2008).

\bibitem{BonnerNossal1981}
R.~Bonner and R.~Nossal, \enquote{Model for laser Doppler measurements of blood flow in tissue.} {\protect\JournalTitle{Applied Optics}} \textbf{20},
  2097--2107 (1981).

\bibitem{Bonner1990}
R.~F. Bonner and R.~Nossal, \enquote{Principles of laser-Doppler flowmetry,} in \enquote{Laser-Doppler Blood Flowmetry,}  (Springer, 1990), pp. 17--45.

\bibitem{Schmitt1992}
J.~Schmitt, A.~Gandjbakhche, and R.~Bonner, \enquote{Use of polarized light to discriminate short-path photons in a multiply scattering medium,}
  {\protect\JournalTitle{Applied Optics}} \textbf{31}, 6535 (1992).

\bibitem{RojasOchoa2004}
L.~F. Rojas-Ochoa, D.~Lacoste, R.~Lenke, P.~Schurtenberger, and F.~Scheffold,
  \enquote{Depolarization of backscattered linearly polarized light,}
  {\protect\JournalTitle{JOSA A}} \textbf{21}, 1799--1804 (2004).

\end{thebibliography}
\end{document}